# Single Channel Speech Enhancement Using Temporal Convolutional Recurrent Neural Networks

Jingdong Li* Hui Zhang*, Xueliang Zhang* and Changliang Li[†]
* College of Computer Science, Inner Mongolian University, Hohhot, China
jingdong.li@mail.imu.edu.cn {cszh, cszxl}@imu.edu.cn
[†] Kingsoft AI Laboratory, Beijing, China
lichangliang@kingsoft.com

*Abstract*—In recent decades, neural network based methods have significantly improved the performace of speech enhancement. Most of them estimate time-frequency (T-F) representation of target speech directly or indirectly, then resynthesize waveform using the estimated T-F representation. In this work, we proposed the temporal convolutional recurrent network (TCRN), an end-to-end model that directly map noisy waveform to clean waveform. The TCRN, which is combined convolution and recurrent neural network, is able to efficiently and effectively leverage short-term ang long-term information. Futuremore, we present the architecture that repeatedly downsample and upsample speech during forward propagation. We show that our model is able to improve the performance of model, compared with existing convolutional recurrent networks. Futuremore, We present several key techniques to stabilize the training process. The experimental results show that our model consistently outperforms existing speech enhancement approaches, in terms of speech intelligibility and quality.

## I. INTRODUCTION

Monaural speech enhancement is the task to extract clean speech from one-microphone noisy signals. The purpose of speech enhancement is to improve speech quality and intelligibility. It is widely and successfully applied in many modern speech applications, such as hearing aids, communication system, automatic speech recognition (ASR) and speaker verification, *etc*[1].

Traditional speech enhancement approaches include spectral subtraction [2], Wiener filtering [3], nonnegative matrix factorization [4] *etc*. These approaches typically rely on the strong assumption that noise has a stationary statistically characteristic. However, there are few noises keep stationary all the time in this complicated world. This makes it hard for traditional methods to achieve satisfactory performance as designed.

To deal with annoying nonstationary noise, deep neural networks (DNNs) [5], [6], [7], [8] are introduced in the speech enhancement, and obtained unprecedented performance. The DNN predicts a label for each frame from a small context window. The limited input makes DNN cannot capture information of a long-term context. The DNN-based methods also perform poorly on unseen speakers. The long short-term memory networks (LSTMs) [9], [10] were introduced into speech enhancement to alleviate the limits of DNN-based methods. Chen *et al*.[10] proposed a four-layer LSTM to deal with speaker generalization of noise-independent speech enhancement. Their experimental results showed that the LSTM model substantially outperforms the DNNs. A more recent study found that a combination of convolution and recurrent network (CRN) [11] leads better performance than LSTM.

Most of the existing approaches are aims to directly or indirectly estimate T-F representations of target speech. They are mainly two groups: "mapping-based" methods and "masking-based" methods. The mapping-based methods directly predict the T-F representations, while the magnitude spectrum of STFT is the most popular choice. The masking-based methods predict a T-F mask at the first stage, then multiply the estimated mask to the T-F features of mixtures to obtain clean features of target speech. In earlier studies, masking-based methods focus on the masks of magnitude spectrum, including ideal binary mask (IBM), ideal ratio mask (IRM) [12], spectral magnitude mask (SMM) [6], phase-sensitive mask (PSM) [9] and so on. Since the performance of magnitude-masking was limited by noisy phase reusing, the complex ideal ratio mask (cIRM) [13] was proposed to improve the performance of speech enhancement. Theoretically, we can get both perfect magnitude and phase using cIRM masking. However, the imaginary part of cIRM exhibits unclear temporal and spectral structure, which is difficult to estimate. It makes cIRM cannot consistently lead to a better performance than other methods.

Recently, some works are developed to use neural networks for speech analysis and synthesis in time domain. Temporal convolutional layers are trained as filterbanks to extract features from waveform to improve the performance of ASR [14], [15], [16]. Compared with hand-crafted mel-filterbank and gamatone-filterbank features, an ASR system jointly trained with trainable filterbanks consistently leads lower word error rate (WER). Sercan *et al*.[17] utilized group convolution networks to synthesis waveform conditioned by magnitude spectrograms. They show that CNN-based methods could generate higher quality speech than signal processing methods, like Griffin-Lim [18]. There are also some works attempted to conduct speech enhancement in time domain. In [19], a CNN-based autoencoder is proposed to conduct speech enhancement in time domain, which outperforms the DNN-based methods in T-F domain. Inspired by these works, we proposed to use temporal convolutional recurrent network (TCRN) to conduct the speech enhancement. Compared with LSTMs and CRN, our proposed model TCRN consistently leads to better speech intelligibility and speech quality.

The rest of the paper organized as follows: section 2 describes the details of the proposed system. Section 3 describes the loss functions used in this study. Section 4 presents the experimental setup and results. Finally, we conclude our work in section 5.

## II. SYSTEM DESCRIPTON

### A. Model Architecture

The proposed temporal convolutional recurrent network (TCRN) is constructed by stacking TCRBs, as showed in figure 1. Compared with the previous methods in time domain [19], the proposed TCRN repeatedly downsample and upsample signals during forward propagation. This makes it possible to use a residual connection between waveforms and waveform's hidden representation. We demonstrate the efficiency of this architecture in the section IV.

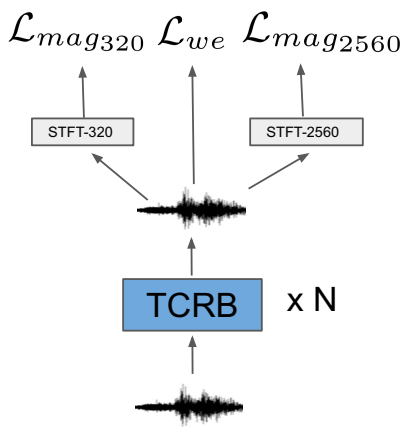

Fig. 1. The proposed TCRN with combined loss.

### B. TCRB Module

We first described the basic building blocks called TCRB, illustrated in Fig. 2. TCRB is consists of a 1-D convolutional layer (Conv), followed by a batch normalization (BN), a LSTM and a 1-D deconvolution (Deconv). TCRB is a powerful module for mapping the noisy waveforms to clean waveforms. The input sample sequence is first convoled with $K$ large 1-D convolutional filters. These filters explicity model the local pattern of the waveform within the receptive field. The convoluton outputs are normalized by BN layer and activated by a Parametric ReLU (PReLU) non-linearity. We use a LSTM to lervege long-term context. The combination of convolution and LSTM can respectively process speech at frame and utterance level. Finally, we stack a deconvolution on top to resynthesize the waveform. Note that we use the symmetry convolution and deconvolution configuration to keep the sigal time-resolution unchanged. There are two residual connections in each TCRB: adding the input of LSTM to the output of LSTM, adding the input of TCRB to the output of Deconv layer. We find that these residual connections are critical for training a deep stacking TCRB architecture. We give the details of each layers in following sections.

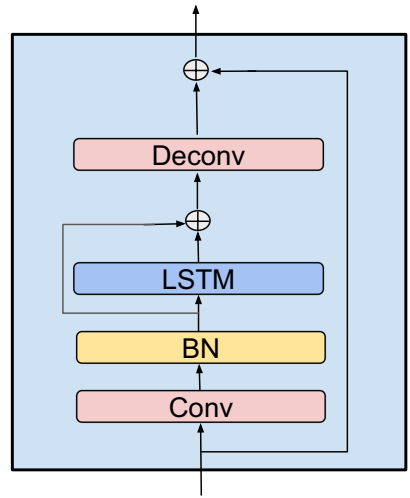

Fig. 2. TCRB: building block of the proposed model

### C. Temporal Convolution

The first component of TCRB is a bank of temporal 1-D convolutional filters, which capture the different local patterns of speech signals. The different feature maps correspond to the different periodic signal components. From a perspective of signal processing, convolutional kernels can be viewed as a group of finite-impulse-response (FIR) filters. Such a layer has the ability to approximate standard filterbanks [20]. Therefore, outputs of the time-convolution is regarded as a hidden T-F representation.

The raw waveform of speech is densely distributed along time. For example, if the speech signals sampled at 16 kHz, then a 20-ms, which is typically used in speech enhancement, will contain 320 samples. This requires the convolution layer has a large receptive field. Some works used deep stacking dilated convolution layers to obtain such a large receptive filed [21], [22]. In our work, we show that simply using large convolutional kernel is also worked in speech enhancement. The similar design is successfully applied in speech recognition [14]. We find that this quite simple and shallow architecture is efficient to model raw waveform in speech enhancement. Furthermore, the outputs of temporal convolutional

A 1-D discrete convolution operator, which convolves signal $F$ with kernel $k$ of size $m$ is defined as:

$$(F * k)(p) = \sum_{s+t=p} F(s)k(t) \qquad (1)$$

where $*$ denotes the convolution operator, and $t \in [-m, m] \cap \mathbb{Z}$. In signal processing, window functions are usually conducted to taper segments of signals. A window pre-processing in time domain helps later subsequent analysis produce more meaningful results. Consequently, we proposed and implemented the *kernel-windowed* 1-D convolution as:

$$(F \circledast k)(p) = \sum_{s+t=p} F(s)W(t)k(t) \quad (2)$$

where $\circledast$ denotes the kernel-windowed convolution operator, and $W$ could be any window functions used in digial signal processing area. In this work, we configured $W$ as symmetric (also called periodic) Hann window, which is commonly used in speech enhancement. During training, the weights of kernel $k(t)$ is updated by gradient descent, but the window $W(t)$ is a group of constant values. We found that *kernel-windowed* convolution could accelerate the convergence of the model in our experiments.

### D. Batch Normalization and LSTM

As mentioned above, the output of 1-D time-convolution is regarded as T-F representations of the raw waveform. For speech enhancement, the T-F features are usually normalized to zero mean and unit variance at each channel. Therefore, a batch normalization layer [23] is introduced to imitate such an operation. After the batch normalization layer, the normalized convolution output is passed to LSTM to get sequential features of target speech.

### E. Temporal Deconvoluton

We utilized a 1-D temporal transposed convolution to upsample the hidden T-F representation back to raw waveforms. The transposed convolution is also called deconvolution. For brevity, we use the name "deconvolution" in the rest of this paper.

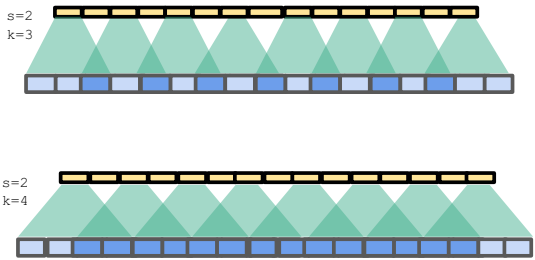

Fig. 3. Up: a 1-D deconv with "uneven overlap", where kernel size $k$ is 3 and stride $s$ is 2. Bottom: a 1-D deconv with "even overlap", where $k$ is 4 and $s$ is 2. We treat the yellow unit in the top layer as the hidden T-F representation, and the blue unit is the results of upsampling operation, where deep blue represent overlapped upsampleing results.

Deconvolution layers allow the model to use every T-F representa=tion vector to generate a longer waveform segment. However, deconvolution can easily have "uneven overlap", causing strange checkerboard pattern of artifacts as illustrated in Fig. 3. Particularly, deconvolution has "uneven overlap" when the kernel size is not divisible by the stride. In this study, we configured the stride as half of the kernel size, so that the output is evenly balanced, to avoid the checkerboard artifacts.

In addition, we also use a window function for the convolutional kernel in deconvolution layers as described in Eq. (2). The deconvolution outputs are divided by the sum-square envelope of a window function, to remove the effects included by windowing observations. We truncate the sum-square of the window function to $[0.1, 1]$ to avoid the numeric problems.

### III. LOSS FUNCTION

In supervised speech separation, loss functions should be correlated with speech quality. We consider the combination of the loss functions defined both in time domain and T-F domain.

### A. Time Domain Loss

We introduced waveform error ($\mathcal{L}_{we}$) as the time domain loss function. For the estimated time domain sigal $\hat{s}$ and the corresponding target signal $s$ with $N$ samples. We defined $\mathcal{L}_{we}$ as:

$$\mathcal{L}_{we} = \frac{1}{N} \sum_{i=1}^{N} (s_i - \hat{s_i})^2$$

The proposed model could be trained with $\mathcal{L}_{we}$ in time domain directly.

### B. T-F Domain Loss

The estimated time domain sigal can also be evaluated in T-F domain. By transforming the estimated output into T-F domain, we compare the it with the target speech signal with the T-F domain loss function. To transform the estimated signal into T-F domain, we use the temporal 1-D convolution described in 2.2 with special and fixed weights to imitate the STFT.

The STFT can be divided into two meaningful parts: the magnitude and the phase in the polar coordinates. Spectral phase is highly unstructured along either time or frequency domain, so fitting errors of the raw phase is also very difficult. We tried it but found that minimizing phase errors makes the training process unstationary. Therefore, we only minimizing the STFT-magnitude loss introduced in [17] as:

$$\mathcal{L}_{mag} = \frac{\| |STFT(s)| - |STFT(\hat{s})| \|_F}{\| |STFT(s)| \|_F} \quad (3)$$

where $\|\cdot\|_F$ is Frobenius norm. We found that the denominator $\| |STFT(s)| \|_F$ reducing the oscillation of $\mathcal{L}_{mag}$ in training.

### C. Loss Combination

In consideration of that the analysis window duration contains different information. We compute the STFT-magnitude losses $\mathcal{L}_{mag320}$ and $\mathcal{L}_{mag2560}$, using a short window (320 points) and a longer window (2560 points), respectively. Due to the loss term $\mathcal{L}_{we}$ and $\mathcal{L}_{mag}$ has different numeric range, we use weight $\alpha$ to balance the importance of all loss term. The combined loss is defined as:

$$\mathcal{L}_{comb} = \mathcal{L}_{we} + \alpha \left( \frac{\mathcal{L}_{mag320} + \mathcal{L}_{mag2560}}{2} \right) \quad (4)$$

TABLE I
MODEL COMPARISONS IN TERMS OF STOI(%) AND PESQ SCORES ON TRAINED NOISE

| SNR | System | Factory | | Babble | | SSN | | Destroyengine | | Destroyerops | | Average | |
|---|---|---|---|---|---|---|---|---|---|---|---|---|---|
| | | STOI | PESQ | STOI | PESQ | STOI | PESQ | STOI | PESQ | STOI | PESQ | STOI | PESQ |
| -5dB | Mixture | 56.6 | 1.40 | 54.3 | 1.30 | 59.4 | 1.39 | 54.7 | 1.27 | 55.2 | 1.41 | 56.0 | 1.36 |
| | LSTM | 76.8 | 1.99 | 73.1 | 1.98 | 71.2 | 1.88 | 69.9 | 1.87 | 71.5 | 1.88 | 72.9 | 1.92 |
| | CRN | 77.2 | 2.14 | 76.6 | 2.00 | 73.3 | 1.97 | 79.3 | 2.07 | 77.9 | 2.03 | 76.4 | 2.04 |
| | TCRN | **78.2** | **2.18** | **77.7** | **2.12** | **76.0** | **2.16** | **82.5** | **2.35** | **82.2** | **2.35** | **79.3** | **2.23** |
| 0dB | Mixture | 68.8 | 1.66 | 66.1 | 1.65 | 69.7 | 1.77 | 66.2 | 1.60 | 67.0 | 1.74 | 67.6 | 1.69 |
| | LSTM | 84.8 | 2.39 | 81.8 | 2.33 | 81.5 | 2.28 | 80.1 | 2.25 | 81.0 | 2.25 | 82.0 | 2.30 |
| | CRN | 84.2 | 2.46 | 84.1 | 2.38 | 82.3 | 2.33 | 86.6 | 2.44 | 84.8 | 2.40 | 84.2 | 2.40 |
| | TCRN | **86.7** | **2.56** | **86.9** | **2.51** | **85.6** | **2.52** | **89.9** | **2.73** | **89.2** | **2.69** | **87.7** | **2.60** |
| 5dB | Mixture | 80.1 | 1.96 | 77.4 | 2.01 | 78.8 | 2.16 | 77.6 | 1.97 | 77.7 | 2.09 | 78.3 | 2.04 |
| | LSTM | 89.0 | 2.67 | 86.6 | 2.60 | 86.6 | 2.59 | 86.1 | 2.57 | 86.5 | 2.56 | 87.0 | 2.60 |
| | CRN | 89.0 | 2.76 | 88.9 | 2.69 | 88.2 | 2.69 | 91.4 | 2.79 | 89.2 | 2.73 | 89.3 | 2.73 |
| | TCRN | **90.9** | **2.82** | **91.4** | **2.82** | **90.5** | **2.82** | **92.5** | **2.94** | **92.2** | **2.93** | **91.5** | **2.87** |

TABLE II
MODEL COMPARISONS IN TERMS OF STOI(%) AND PESQ SCORES ON UNTRAINED NOISE

| SNR | System | Factory2 | | M109 | | Cafe | | Street | | Pedestrian | | Average | |
|---|---|---|---|---|---|---|---|---|---|---|---|---|---|
| | | STOI | PESQ | STOI | PESQ | STOI | PESQ | STOI | PESQ | STOI | PESQ | STOI | PESQ |
| -5dB | Mixture | 65.2 | 1.56 | 68.2 | 1.70 | 55.2 | 1.34 | 67.2 | 1.59 | 61.1 | 1.55 | 63.4 | 1.55 |
| | LSTM | 78.4 | 2.18 | 79.5 | 2.29 | 65.4 | 1.69 | 78.1 | 2.21 | 71.7 | 1.92 | 74.7 | 2.06 |
| | CRN | 81.1 | 2.27 | 81.2 | 2.32 | 65.9 | 1.74 | 80.5 | 2.26 | 73.2 | 1.98 | 76.3 | 2.11 |
| | TCRN | **83.9** | **2.41** | **83.5** | **2.39** | **69.0** | **1.80** | **83.0** | **2.34** | **76.7** | **2.05** | **79.2** | **2.20** |
| 0dB | Mixture | 75.3 | 1.93 | 77.7 | 2.07 | 66.3 | 1.66 | 75.4 | 1.95 | 72.5 | 1.89 | 73.4 | 1.90 |
| | LSTM | 85.1 | 2.53 | 85.4 | 2.59 | 77.8 | 2.13 | 84.3 | 2.53 | 81.8 | 2.31 | 82.9 | 2.42 |
| | CRN | 87.2 | 2.61 | 87.4 | 2.67 | 78.2 | 2.15 | 86.3 | 2.61 | 83.2 | 2.39 | 84.5 | 2.48 |
| | TCRN | **90.2** | **2.76** | **90.0** | **2.73** | **82.6** | **2.28** | **89.6** | **2.70** | **86.6** | **2.47** | **87.8** | **2.59** |
| 5dB | Mixture | 84.0 | 2.29 | 85.2 | 2.41 | 77.0 | 2.03 | 83.1 | 2.34 | 81.9 | 2.23 | 82.2 | 2.26 |
| | LSTM | 88.8 | 2.79 | 88.8 | 2.82 | 85.4 | 2.51 | 88.1 | 2.78 | 87.3 | 2.62 | 87.7 | 2.70 |
| | CRN | 91.0 | 2.92 | 91.3 | 2.96 | 86.6 | 2.56 | 90.5 | 2.92 | 89.2 | 2.72 | 89.7 | 2.82 |
| | TCRN | **93.0** | **3.01** | **92.9** | **3.00** | **89.8** | **2.67** | **92.7** | **2.97** | **91.2** | **2.80** | **91.9** | **2.89** |

## IV. EXPERIMENTS

### A. Experimental Setup

In our experiments, we evaluate the models on the TIMIT dataset [24]. 2000 utterances from TIMIT training set are randomly chosen as the training set. All 192 utterances from the TIMIT core test set are used for test. Five types of noise are used for training: babble, factory1, destroyerengine, destroyerops noise from NOISEX-92 dataset [25] and a speech-shaped noise (SSN). These five types of noise are also used in the noise-depend evaluation. For noise-independent evaluation, we use 5 different noises from different datasets: pedestrian, cafe, street noises from CHiME-4 [26] dataset and factory2, tank (m109) from NOISEX-92. These noises are all highly non-stationary, which makes speech enhancement be a challenging task. The training set are formed by mixing all the speech and the noises at {-5, 0} dB signal-to-noise ratio (SNR). Each utterance in the training set is repeatedly used 5 times with mixed with different segments of noises, producing $2000(utterances) \times 5(noise) \times 2(SNR) \times 5(repeat) = 100000$ training mixtures in total. The test mixtures are constructed by mixing random cuts from noises with test utterances at {-5, 0, 5} dB SNR, which contains one unseen SNR (5 dB) in training. All signals are resampled to 16 kHz before mixing.

We use the Adam [27] optimizer with learning rate 0.001 to minimize the combined loss. We train the models using a batch size of 32. Within a mini-batch, all sequences are zero-padded to the length divisible by 160. In our experiments, the $\alpha$ in combined loss is set to 0.1.

### B. Baselines

We use TCRN with four TCRB layers to compare with LSTM and CRN baselines. The LSTM baseline has 161, 1024, 1024, 1024, and 1024, 161 units, respectively. For CRN, we configured the network using the same hyper-parameters described in [11], that are well tuned. Both baseline models are mapping from 161-D magnitude spectrum of noisy speech to 161-D magnitude spectrum of target speech. And the phase of noisy speech is used to reconstruct the waveforms. In addition, the proposed method TCRN and baselines are all causal systems, do not use future information.

### C. Experimental Results

In this study, speech enhancement performance is evaluated in terms of short-term object intelligibility (STOI) and perceptual evaluation of speech quality (PESQ) [28]. For both metrics, a higher score means better performance.

Tab. I and Tab. II present STOI and PESQ scores of unprocessed and processed signals for trained noise and untrained noise, respectively. In each case, the best result is highlighted by boldface. As shown in Tab. I and II. The proposed TCRN significantly outperforms the LSTM baseline with a large margin. And the propose TCRN also leads to consistently better metrics than CRN. Comparing the results in Tab. II, we can find that TCRN has better noise generalization ability than baselines.

## V. CONCLUSIONS

In this study, we proposed a temporal convolutional recurrent network to deal with speech enhancement in time domain. The proposed TCRN is consistently superior to LSTM and CRN in the T-F domain. We believe that the proposed model lays a sound foundation for supervised speech enhancement in time domain. Future research includes exploring the proposed TRCN for speaker separation or music source separation in time domain.